\documentclass[useAMS,usenatbib,fleqn]{mn2e}
\usepackage{amsmath,amssymb, aas_macros,times}
\usepackage{graphicx}
\graphicspath{ {figures/} }
\usepackage[usenames,dvipsnames]{xcolor}
\usepackage{color}
\newcommand{\beq}{\begin{equation}}
\newcommand{\eeq}{\end{equation}}
\newcommand{\beqarray}{\begin{eqnarray}}
\newcommand{\eeqarray}{\end{eqnarray}}
\usepackage{float}

\newcommand{\be}{\begin{equation}}
\newcommand{\ee}{\end{equation}}
\newcommand{\bse}{\begin{subequations}}
\newcommand{\ese}{\end{subequations}}
\newcommand{\bary}{\begin{eqnarray}}
\newcommand{\eary}{\end{eqnarray}}
\newcommand{\bwt}{\begin{widetext}}
\newcommand{\ewt}{\end{widetext}}

\begin{document}
\title[SED modelling of Cen A]{Very high-energy gamma-ray signature of ultrahigh-energy cosmic-ray acceleration in Centaurus A}

\author[Joshi et al.]{
Jagdish C. Joshi$^{1}$\thanks{jjagdish@uj.ac.za},
Luis Salvador Miranda $^{1}$,
Soebur Razzaque$^{1}$\thanks{srazzaque@uj.ac.za} and
Lili Yang$^{2,1}$\\
$^{1}$ Department of Physics, University of Johannesburg, PO Box 524, Auckland Park 2006, South Africa\\
$^{2}$ Center for Astrophysics and Cosmology, University of Nova Gorica, Vipavska 13, Nova Gorica, Slovenia
}
\bibliographystyle{mn2e}


\maketitle

\begin{abstract}
The association of at least a dozen ultrahigh-energy cosmic-ray (UHECR) events with energy $\gtrsim 55$ EeV detected
by the Pierre Auger Observatory (PAO) from the direction of Centaurus-A, the nearest radio galaxy, supports the scenario
of UHECR acceleration in the jets of radio galaxies. In this work, we model radio to very high energy (VHE,$\gtrsim 100$ \rm GeV) $\gamma$-ray emission
from Cen A, including GeV hardness detected by Fermi-LAT and TeV emission
detected by HESS. We consider two scenarios: (i) Two zone synchrotron self-Compton (SSC) and external-Compton (EC)
models, (ii) Two zone SSC, EC and photo-hadronic emission from cosmic ray interactions. The GeV hardness observed by Fermi-LAT can be explained using these two scenarios, where zone 2 EC emission is very important.
Hadronic emission in scenario (ii) can explain VHE data with the same spectral slope as obtained through fitting UHECRs from Cen A. The peak luminosity in cosmic ray proton at 1 TeV, to explain the VHE $\gamma$-ray data
is $\approx 2.5 \times 10^{46}$ erg/s. The bolometric luminosity in cosmic ray protons is
consistent with the luminosity required to explain the origin of 13 UHECR signal events that are correlated with Cen A.

\end{abstract}
 
\date{\today}
\maketitle

\section{Introduction}
Centaurus A (Cen A) or NGC 5128 is a radio galaxy (elliptical shape) of FR-I type \citep{fanarof_riley} with inner X-ray jet of order 1 kpc \citep{israel_fp}, located at a distance of 3.7 Mpc \citep{dist_to_cenA_2007}. This galaxy has two giant radio lobes with $10^{\circ}$ extension
along the north-south direction in the sky \citep{feain_radiolobes}. The substructures in Cen A, like inner lobes and jets have been studied in radio wavelength and X-ray band. The radio observations detected a
subluminal motion $(v\sim 0.5c)$ in the 100 pc length of the jet \citep{hardcastle_jet_randx}. This analysis assumes jet-counterjet symmetry in Cen A, and  estimates that the jet makes an angle $\theta_{\rm ob} = 15^{\circ}$ with the line of sight (l.o.s.) of the observer.
Another study of Cen A uses the bright jet subluminal motion $(v\sim 0.1c)$ and constraints that the jet axis can be in between $50^{\circ}-80^{\circ}$ w.r.t. the l.o.s. of the observer \citep{higher_angle_tingay}.
In this work, we consider $\theta_{\rm ob} =30^{\circ}$, which has been used in the SSC modelling of Cen A \citep{coreview_2010_cena}. The central region activity of Cen A is powered by a supermassive black hole whose mass has been estimated using the gas kinematics around Cen A \citep{smbh_cena_07}, and
matches the more recent update on its mass of value  $(5.5 \pm 3.0) \times 10^7 M_{\odot}$ \citep{smbh_cena_09} using the stellar
kinematics studies. This provides the Eddington luminosity for Cen A of value $L_{\rm Edd} = (0.7 \pm 0.4) \times 10^{46}$ erg/s. 

The detection of $\gamma$-rays from Cen A started in $1970s$ \citep{1975_cena_gray,nucleargrays} and continued until $1990s$ \citep{sreekumar_99_cena}. In the last decade, more data from this object have been detected
in MeV-GeV $\gamma$-rays by the Fermi satellite \citep{coreview_2010_cena} and in the VHE range by ground based Cherenkov telescopes \citep{rowell_teV,aharonian_hesstev1}. Using 4 years of Fermi-LAT data of Cen A,
\cite{sahakyan_13_seccomp_cena} discovered that the $\gamma$-ray spectrum becomes harder above approximately 4 GeV. In 2007, the PAO correlated two cosmic ray events within $3.1^{\circ}$ of Cen A \citep{pao_2007_2evnt}. Until 2015, in 10 years of PAO data 13 events above 55 EeV (from the central $18^{\circ}$ region of Cen A) are correlated to this source, where
3.2 events are expected from the isotropic background \citep{uhecr_event_CenA}. This correlation has been increased to 19 events in 2017, where 6 events are expected from the isotropic background \citep{pao_2017_icrc}.

\noindent Observations of Cen A, before the discovery of GeV hardness in 2013 \citep{sahakyan_13_seccomp_cena}, are well explained using one zone SSC models \citep{firstSSC_paper,coreview_2010_cena} and in multi-zone SSC models \citep{2008_lenain},
if the HESS observation is not taken into account \citep{aharonian_hesstev1}. \cite{RoustazadehnBottcher2010} considered $\gamma-\gamma$ interactions in the internal jet and associated cascade emission to explain the GeV emission.
In order to model the HESS data it was assumed that protons, Fe nuclei are accelerated in the inner jet of Cen A and their interaction with synchrotron photons can produce VHE emission \citep{sahu_cena, pd_Fe_cena}.
More recent models explain the GeV hardness and VHE spectrum by considering the photo-disintegration of Fe-nuclei in the jet of Cen A \citep{kundu_gupta_cenA}. Similarly a revised one zone SSC model combined with
photo-hadronic interaction in the jet can explain the broad spectral energy distribution (SED) of Cen A \citep{petropolou,fraija_muon}.

In this work, we introduce two zone SSC+EC and SSC+EC+photo-hadronic models for Cen A. The SED of Cen A has been explained in two scenarios:(i) Pure leptonic scenario, where SSC is important in zone 1 and EC in zone 2;
(ii) Lepto-hadronic scenario, where SSC and photo-hadronic interactions are important in zone 1 and EC in zone 2. In sections 2 and 3 we discuss leptonic and photo-hadronic models, respectively,
following with $\gamma - \gamma$ pair production opacities in the emission zones. In section 4, we discuss spectral fitting of UHECR events from Cen A and calculate power from the spectrum along with deflections of UHECRs in magnetic fields. We summarise our work in section 5.

\section{Leptonic emission models}

The SED peaks of Cen A, at $10^{-1}$ eV (infrared), and at 170 keV (soft $\gamma$ rays) are reasonably explained by the leptonic models \citep{firstSSC_paper, coreview_2010_cena, petropolou}.
We use a publicly available code by \cite{krawczynski} to calculate SSC and EC radiative losses. In this code the input parameters of the model are the redshift
$z$ of Cen A, the size of the emitting blob $R_{\rm b}$, energy density of electrons in this blob $w_e$, magnetic field $B$, the jet angle w.r.t. the observer $\theta_{\rm ob}$ and a spectrum of electrons with a break energy. We
estimate the spectrum of thermal photons from the disk, EC with the disk photons and the SSC process in the jet of Cen A.
The SSC parameters $B, R_{\rm b}, \theta_{\rm ob}$, bulk-Lorentz factor of the jet ($\Gamma$) and Doppler factor ($\delta_D = [\Gamma (1-\beta \cos(\theta_{\rm ob}))]^{-1}$), where $\beta$ is the speed of the relativistic jet, are listed in Table \ref{tab:sumpar}. The injected spectrum of electrons follows
a broken power law with spectral index $p1$ from minimum electron energy $E_{e,\rm min}$ to the break energy $E_{e,\rm br}$ and $p2$ from $E_{e,\rm br}$ to the maximum electron energy $E_{e,\rm max}$, as

\begin{align}
\begin{array}{ll}
\frac{dN_e}{dE_e} = N_0 \times \begin{cases}
\left(\frac {E_e}{E_{e,\rm br}} \right)^{-p1}; & E_{e} \leq E_{e,\rm br}\\
\left(\frac {E_e}{E_{e,\rm br}}\right)^{-p2}; & E_{e} > E_{e,\rm br}\
\end{cases}
\end{array}
\label{add_spec}
\end{align}
The $\gamma$-ray flux in the SSC model is produced due to interaction of theses electrons with target photons available in the jet of Cen A. Similarly in the EC model $\gamma$-rays are produced by the interaction of high energy electrons with the
target radiation field available outside the jet, which is from the accretion disk.



\begin{table*}
\centering
\vskip 0.5cm
\caption{Parameters used for the leptonic and hadronic modelling of Cen A. The Schwarzschild radius $R_{\rm s}$ for Cen A with black hole mass of $5.5 \times 10^7 M_{\odot}$ is $1.65 \times 10^{13}$ cm.
Zone 1 is for SSC or SSC+photo-hadronic emission, respectively, in the leptonic or lepto-hadronic case. Zones 2a and 2b represents EC scenario in leptonic and lepto-hadronic cases, respectively.}
\vskip 0.25cm
\begin{tabular}{|l|l|l|l|l|}
\hline
Physical parameters & Leptonic/lepto-hadronic (zone 1) & Leptonic (zone 2a) & Leptonic (zone 2b) \\
\hline
 Magnetic field $(B)$ & $ 2.5  $ G &$ 0.02  $ G  &$ 0.02  $ G    \\
 Blob radius ($R_b$) &  $7 \times 10^{15}$ cm  & $ 9.2 \times 10^{15}$ cm & $ 9.2 \times 10^{15}$ cm \\
 Lorentz boost factor ($\Gamma$) & 7.0 & 2.0 & 2.0 \\
 Angle between jet and l.o.s. ($\theta_{\rm ob}$) &  $30^{\circ}$ &  $30^{\circ}$&  $30^{\circ}$ \\
 Doppler factor ($\delta_D$)& 1.0 & 2.0 & 2.0 \\

 \hline
 Electron spectrum:\\
 ($E_{e,\rm min},E_{e,\rm br}, E_{e,\rm max}$) & ($10^{8.3},10^{8.7},10^{12}$) eV& ($10^{8.0},10^{9.25},10^{13.2}$) eV & ($10^{8.0},10^{8.35},10^{11}$) eV\\
 ($p_1,p_2$) & ($1.85,4.3$) & ($1.8,3.1$)& ($1.8,2.28$)\\
Jet-frame energy density of electrons ($\rm erg/cm^3$) & 0.37  & 0.03  &  0.005 \\
Emitting blob location along the jet axis & 3030$\times  R_{\rm s}$    & 4030$\times R_{\rm s}$  & 4030$\times R_{\rm s}$\\ 
\hline
 Proton spectrum:\\
 ($E_{p,\rm min},E_{p,\rm br}, E_{p,\rm max}$)  &($1,10^3,10^{11}$) GeV  & &\\
 ($\alpha_1,\alpha_2$) &(2.0, 2.5)  & & \\
\hline
\end{tabular}
\label{tab:sumpar}
\end{table*}

\begin{figure*}
  \includegraphics[width=11.5cm,height=8cm]{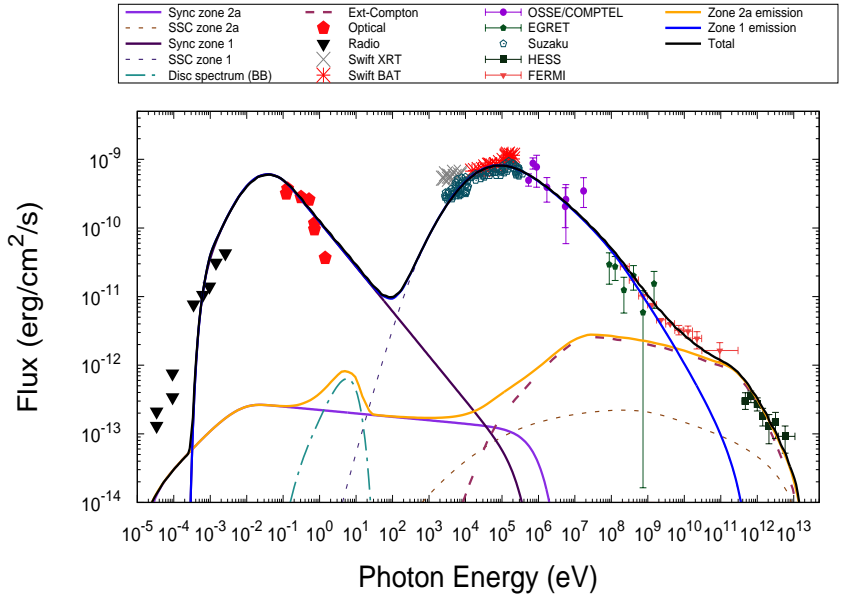}
  \caption{The SED of Cen A, in two zone leptonic scenario. The data points from radio to GeV energy are taken from \citep{coreview_2010_cena} and references therein. The recent HESS observational data in 213 hours and Fermi-LAT data in 8 years
  have been used in this plot \citep{hessferminew}.
  The fluxes from zone 1 are dominant in synchrotron and SSC emission. The zone 2 emission dominates in the VHE energy range and fits the Fermi-GeV hardness as well as the HESS data. The emission mechanism in zone 2 is EC.
  The parameters of this model are shown in Table \ref{tab:sumpar}. The excess radio emission compared to the model is likely due to a larger radio-emitting region than the jet structure we have used.
  In general, the blob scenario of AGNs underpredict radio data \citep{krawczynski,coreview_2010_cena,petropolou}.}
  \label{fig:purelep}
\end{figure*}


\section{Photo-hadronic model}
Cen A is a probable source of UHECRs as seen in the PAO association of $\gtrsim 55$ EeV events towards its location. Acceleration of cosmic ray protons upto 50 EeV can occur via shear acceleration in the jet of Cen A \citep{rieger_and_cray_cena}
while diffusive shock acceleration in the jet of Cen A can accelerate protons up to 100 EeV \citep{accn_ppr_by_honda}. If $L$ is the apparent isotropic luminosity of Cen A at a radius $R$, and if a fraction $\epsilon_B$ of this luminosity goes into magnetic field and a fraction $\epsilon_e= L_{\gamma}/L$ into electrons, then the maximum energy of
the particles with atomic number $Z$, accelerated in Cen A, \citep{dermer_razzaque09,dermer_razzaque_accnppr} is given by,
\begin{equation}
 E_{\rm max} \cong 2 \times 10^{20} Z \frac{\sqrt{(\epsilon_B/\epsilon_e)\beta L/10^{46}~\rm{erg/s}}}{\Gamma} \rm eV 
\end{equation}

\noindent The maximum energy of cosmic rays in Cen A can also be derived by Fermi-acceleration in colliding shells using the apparent source power $L$ and the bulk Lorentz factor of the shells $\Gamma, \Gamma_a$ \citep{dermer_razzaque09},

\begin{equation}
 E_{\rm max} \le 2.4 \times 10^{20} Z \frac{\Gamma}{\Gamma_a^2} \sqrt{\epsilon_{B}L/10^{46}~\rm{erg/s}}~ \rm eV 
\end{equation}

\noindent The emission from Cen A is reported in quiescent state and the photo-hadronic models to explain this scenario can be adopted from the earlier models \citep{Sahu:2015tua,Sahu:2016mww}.
The energy of a photon for Cen A, in comoving frame (denoted by $\prime$) and in the observer frame (no superscript)
are related by $E_{\gamma}=\delta_D E_{\gamma}^{\prime}/(1+z)$.
The kinematical condition $E_\gamma \epsilon_\gamma \simeq 0.032~\frac{ {\delta_D}^2}{(1+z)^2} ~{\rm GeV}^2$ relates the photons of energy $E_{\gamma}$ produced via pion decay during the $p-\gamma$ interactions, to the target photons
of energy $\epsilon_{\gamma}$. We use a broken power-law for proton distribution but here, for mathematical expressions, we have shown the formalism for an individual segment of the power law. For the protons with energy $E_p$ it is given as,

\be
\frac{dN_p}{dE_p} = C_{p} E_p^{-\alpha},
\label{powerlaw}
\ee

\noindent with a spectral index $\alpha \ge 2$. These protons interact with the target photons and create neutral pions which decay to $\gamma$-rays. The number of $\pi^0$-decay photons at a given energy is proportional
to both the number of high energy protons and the density of the SSC background photons in the jet, i.e. $N(E_{\gamma})=C_{\gamma} N_p(E_p) n'_{\gamma}(\epsilon_\gamma)$. The $\gamma$-ray flux from the $\pi^0$ decay is then
given by ($E_{p}=10 \Gamma E_{\gamma}/{\delta_D}$),

\be
F_{\gamma}(E_{\gamma}) \equiv E^2_{\gamma} \frac{dN(E_\gamma)}{dE_\gamma} 
= C_{p} C_{\gamma} (10 \Gamma/\delta_D)^{1-\alpha} n'_{\gamma}(\epsilon_\gamma)E_{\gamma}^{2-\alpha}
\label{eq:sebas}
\ee

\noindent In terms of the SSC photon energy and its luminosity, the photon number density $n_{\gamma}^{'}$ is expressed as
\be
n_{\gamma}^{'}(\epsilon_{\gamma})=\eta \frac{L_{\gamma,\rm SSC}(1+z)}{{\delta_D}^{2+\kappa} 4\pi R_{\rm b}^{'2} c \epsilon_{\gamma}},
\label{eq:irma}
\ee
where $\eta \sim1.0$, $\kappa$ describes whether the jet is continuous ($\kappa=0$) or discrete ($\kappa=1$) \citep{sahu_cena}. 
The SSC photon luminosity is expressed in terms of the observed flux
$\Phi_{\rm SSC}(\epsilon_{\gamma})=\epsilon_{\gamma}^{2}dN_{\gamma}/d\epsilon_{\gamma}$ and is given by,
\be
L_{\gamma,\rm SSC}=\frac{4\pi d_{L}^{2}\Phi_{\rm SSC}(\epsilon_{\gamma})}{(1+z)^2}
\label{eq:irma2}
\ee

\noindent The $\Phi_{\rm SSC}$ is calculated using the leptonic model. 
Taking the ratio of fluxes at two energies, using Eq. (\ref{eq:sebas}) and kinematic condition, the flux for an energy $E_{\gamma}$ takes the form \citep{Sahu:2016mww},

\be
F(E_{\gamma})=A_{\gamma} \Phi_{\rm SSC}(\epsilon_{\gamma} )\left (
  \frac{E_{\gamma}}{\rm GeV}  \right )^{-\alpha+3}
\label{eq:modifiedsed}
\ee

\noindent The optical depth of the $\Delta$-resonance process with cross section $\sigma_{\Delta}=5 \times 10^{-28} \rm cm^2$ is $\tau_{p\gamma}=n'_{\gamma} \sigma_{\Delta} R'_{\rm b}$.
In this picture, one out of $\tau^{-1}_{p\gamma}$ protons, interacts with the SSC background to produce photons and neutrinos . So the fluxes of the TeV photons and the
Fermi accelerated high energy protons $F_p$, are related through,
\be
F_p(E_p)=5\times\frac{3}{2} \frac{1}{\tau_{p\gamma}(E_p)}
F_{\gamma}(E_{\gamma}),
\label{eq:protonflux}
\ee
where the factor 5 corresponds to $\approx 20\%$ of the proton energy taken by each pion \citep{waxmann_fraction_ener} and 3/2 is due to the 2/3 probability of $\Delta$-resonance decaying to $p\pi^0$. The $\gamma$-ray flux from photo-hadronic interactions
is shown in Fig. \ref{fig:photohadr}. We used $\alpha=2$ below 1 TeV and $\alpha=2.5$ above 1 TeV in Eq. (\ref{powerlaw}). 

\subsection{$\gamma-\gamma$ opacities in emission zones}
\begin{figure}
\includegraphics[width=8. cm,height=11cm,keepaspectratio,angle =0]{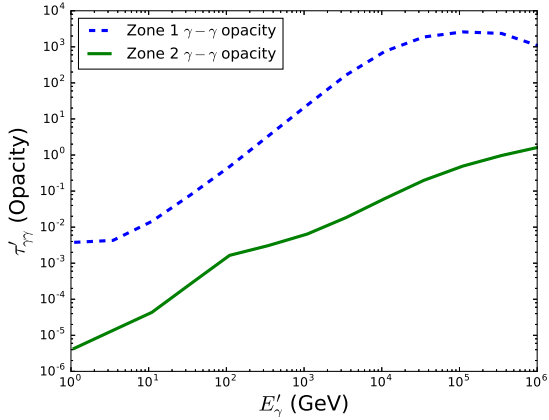}
\caption{The $\gamma-\gamma$ opacity is shown for the GeV-PeV photons in the target radiation field. In zone 1, $p-\gamma$ interaction produces GeV-PeV photons but they undergo severe attenuation losses in the target radiation field
of this zone. The level of zone 2 opacity is negligible for $\lesssim 1$ PeV.}
\label{fig:opaqplot}
\end{figure}

The accelerated protons in the jet of Cen A, interacts with the target photons available in zone 1 primarily and negligibly in zone 2.
Before escaping their respective zones, VHE photons interact with the SSC target radiation field. For head on collision, the pair production condition is $ E_{\gamma}^{\prime} \epsilon_{\gamma}^{\prime} \ge 2(m_ec^2)^2$.
We estimate the opacities of zone 1 and zone 2 using delta function approximation for the $\gamma-\gamma$ cross section.
In this approximation the $\gamma-\gamma$ opacity in the comoving frame is given by,
\begin{equation}
 \tau_{\gamma\gamma}^{\prime}(E_{\gamma}^{\prime}) = \frac{\sigma_{\rm T}}{5} R_{\rm b}^{\prime} n_{\gamma}^{\prime}\left(\frac{2(2m_ec^2)^2}{E_{\gamma}^{\prime}}\right)
 \label{ggo}
\end{equation}
here $\sigma_{\rm T} = 6.6524 \times 10^{-25} \rm cm^{2}$ is the Thomson cross section, $E_{\gamma}^{\prime}$ is the higher energy photon and $\epsilon_{\gamma}^{\prime}$ is the energy of target photon in zone 1 or 2.

\begin{figure*}
  \includegraphics[width=11.5cm,height=8cm]{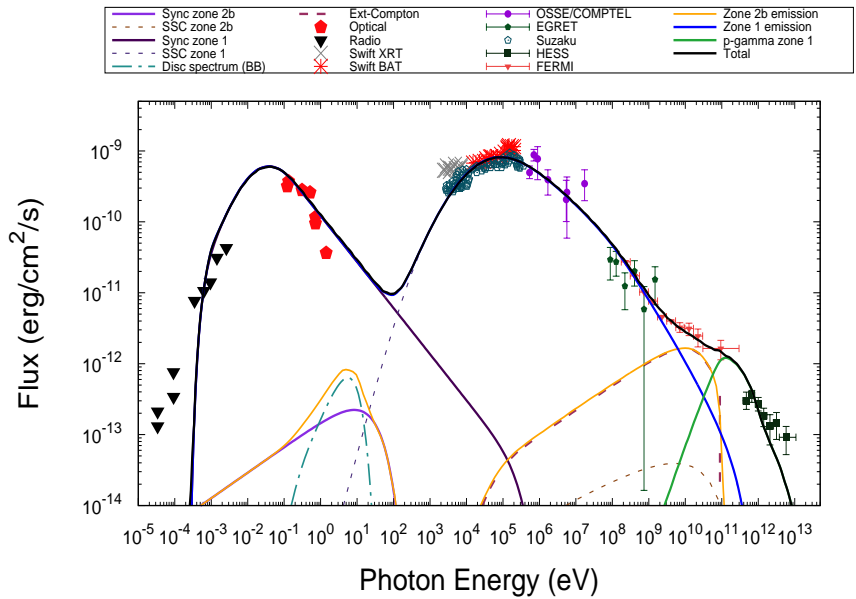}
  \caption{The SED of Cen A, in two zone leptonic and photo-hadronic scenario. The SSC emission from zone 1 is similar to the pure leptonic model in Fig. \ref{fig:purelep}. We calculate the VHE $\gamma$-ray flux using the photo-hadronic interaction in zone 1. The attenuation impacts on these photons due
  to zone 1 target photons (see Fig. \ref{fig:opaqplot}), suppress the VHE spectrum. Here the emission from zone 2 in the EC scenario has been tuned to produce the GeV hardness observed in the spectrum of Cen A.}
  \label{fig:photohadr}
\end{figure*}

The variation of opacity for a photon of energy $E_{\gamma}^{\prime}$ in zone 1 is shown in Fig. \ref{fig:opaqplot}. We find that $\tau_{\gamma \gamma}^{\prime}>1$ above 200 GeV and increases with
energy of pion decay photons. The opacity in zone 2 is significantly lower compared to the zone 1, which means $\gamma$-rays easily escape from this zone.
In zone 1 the $p-\gamma$ interaction efficiency is higher due to higher flux of target photons around 170 keV. VHE photons in Fig. \ref{fig:photohadr} therefore are attenuated significantly in zone 1 and negligibly in zone 2.
The opacity effect are estimated using the slab approximation,
which attenuates the flux by a factor $[(1.0-\rm exp(-\tau_{\gamma \gamma}^{\prime}))/\tau_{\gamma \gamma}^{\prime}]$ \citep{2009_astroparticlephy_incompact_obj}.

Another opacity effect may occur in the broad-line region (BLR) of Cen A, where target photons at infrared wavelength ($1.5~ \mu m ~\rm{or}~ 0.8~eV$) with luminosity $10^{41}$ erg/s, and at shorter wavelengths
($0.337 ~\mu m ~\rm{or}~ 3.6~ eV$), where the luminosity is less than $10^{36}$ erg/s \citep{marconi}. The BLR radius of Cen A, $R_{\rm BLR} = 10^{17}\left(L_{\rm disk}/10^{45}\rm{erg/s}\right)^{0.5}$ cm \citep{blrradiush_isellini},
where $L_{\rm disk}= 10 L_{\rm BLR}$, is $3.2 \times 10^{15}$ cm $\approx$ 0.001 pc, assuming $L_{\rm disk} =10^{42}$ erg/s. This is below the location of two emitting blobs located at 0.016 pc and 0.022 pc respectively. In our model, VHE photons are not attenuated in the BLR of Cen A.

\section{UHECR events from Cen A}
\label{sec:photo}
In the 10 years dataset of PAO, some events are correlated with Cen A, using the exposure of the detector $\Xi = 66,000$ km$^2$ sr yr \citep{uhecr_event_CenA}. A correction factor
$\omega(\delta_s) \simeq 0.64$ should be used for the Auger exposure for Cen A, which is $\Xi \omega(\delta_s)/ \Omega_{80} \simeq(66,000 \times 0.64/5.19)$ km$^2$ yr, where $\Omega_{80}$ is the acceptance solid angle used in PAO.
The cosmic ray spectrum $dN_p/dE_p=N_0 (E_p/E_0)^{-\alpha}$ with normalization $N_0 \rm (1/km^2/yr/EeV)$, $E_0 = 1\rm EeV$ and spectral index $\alpha$ can be explored to get some constraints from Auger's observations. 
If $N$ is the number of events discovered by the PAO then we can write \citep{cuoco_cenA_neutr},

\begin{equation}
 N = \frac{\Xi \omega(\delta_s) N_0E_0}{\Omega_{80} (\alpha-1)} \left (\frac{E_{\rm th}}{E_0}\right)^{1-\alpha}
\end{equation}

\noindent where $E_{\rm th}$ = 55 EeV is the threshold energy of cosmic rays for detection at PAO \citep{uhecr_event_CenA}. The cosmic ray spectrum for N cosmic ray events,

\begin{equation}
\frac{dN_p}{dE_p} = 1.2 \times 10^{-4} N (\alpha-1) \left( \frac{E_{\rm th}}{E_0}\right)^{\alpha-1} \left( \frac{E_P}{E_0}\right)^{-\alpha}
\end{equation}
Using the 13 event data set for Cen A \citep{uhecr_event_CenA} we calculate the spectral index to be $2.55 \pm 0.2$ with reduced $\chi^2$=0.7 for 3 degrees of freedom.
We also fit the HESS $\gamma$-ray observation of Cen A using the photo-hadronic model and
the spectral index was taken as 2.5 above 1 TeV. The peak lumonisity in cosmic rays at 1 TeV to explain HESS observations is $\approx 2.5 \times 10^{46}$ erg/s.
The bolometric luminosity for cosmic ray protons in the range 100 GeV-100 EeV is approximately $10^{47}$ erg/s, which is an order of magnitude higher than $L_{\rm Edd}$. Such super-Eddington luminosity is often required for blazar cosmic-ray
emission \citep{razzaque_lum_ppr}. For 10 UHECR events above 55 EeV the luminosity is $1.6 \times 10^{39}$ erg/s, while the total luminosity
in cosmic rays above 55 EeV is approximately $10^{42}$ erg/s for the spectrum required to explain VHE $\gamma$-ray data.

Cosmic rays emitted from Cen A will go through magnetic deflection in the Galactic and intergalactic magnetic field. In the intergalactic medium the deflection can be neglected while in the Galactic case if an UHECR event of energy $E$,
enters our Galaxy at latitude $b$ then its deflection $\theta_{\rm dfl,MW}$ in the magnetic field ($B$) of our Galaxy can be estimated using \citep{dermer_razzaque09},
\begin{equation}
 \theta_{\rm dfl,MW} \lesssim 1^{\circ}\frac{Z h_{\rm md}(\rm{kpc})B(\mu G) }{{\rm sin} b (E/60 \rm EeV)}
\end{equation}

\noindent Where $h_{\rm md}$ is the height of the Galactic disk. If we assume, the emitted cosmic ray composition favours cosmic ray protons then their deflection in the Galactic field can be assumed within $1^{\circ}$.
The deflection in the intergalactic medium is negligible for the distance of Cen A \citep{dermer_razzaque09}.

\label{sec: deflection}

\section{Summary and discussion}
\label{sec:conclusion}

The SSC model parameters $\Gamma, \theta_{\rm ob}, \delta_D$ used in our model for zone 1, are the same as used in \cite{coreview_2010_cena}, while we change the magnetic field $B$ and the blob size $R_b$.
The parameters of these two models are shown in Table \ref{tab:sumpar}. We show that VHE $\gamma$-ray emission gets attenuated severely in zone 1 of Cen A for $\delta_D=1.0$, which was neglected in earlier works \citep{sahu_cena,pd_Fe_cena}. One zone photo-hadronic model has been discussed
for $\delta_D=1.0$, which explains the GeV hardness and $\delta_D=2.0$, which allows VHE $\gamma$-ray to escape from Cen A \citep{petropolou}. In this work, we consider $\delta_D=1.0$ case, where VHE $\gamma$-ray
attenuation in zone 1 are calculated to explain SED of Cen A in two zone pure leptonic model and lepto-hadronic model. The GeV hardness in our model has been addressed using a second zone with $\delta_D =2.0$ where the EC process is effective. 

We find that both pure leptonic and lepto-hadronic scenarios in our two-zone model can satisfactorily explain the SED of Cen A. The association of UHECR events with Cen A
naturally brings into question of electromagnetic signature of these energetic particles. In our lepto-hadronic model, protons with the same spectral shape by fitting UHECR data from Cen A \citep{uhecr_event_CenA} can
account for observed VHE $\gamma$-ray emission through photo-hadronic interactions. The luminosity in protons above 55 EeV is $10^{42}$ erg/s for the spectrum required to fit VHE $\gamma$-ray data.
This is roughly a factor $6\times 10^2$ higher than the luminosity in UHECRs from Cen A above 55 EeV.
This factor could arise from the fact that the jet of Cen A is $30^\circ$ off axis from our line of sight and only a fraction of the particles in the jet can reach us,
which are significantly deflected by the magnetic field in the large-scale jet and lobes to our line of sight \citep{dermer_razzaque09}.

Cen A, the nearest radio galaxy, is a test bed for applying our knowledge of particle acceleration to ultrahigh energies, study and model their interactions to interpret data.
It will continue to be an intriguing source in foreseeable future. 
\section*{Acknowledgements}
J.C.J. acknowledges support from a GES fellowship at the University of Johannesburg (UJ) while S.R. acknowledges support by grants from UJ and from the National Research Foundation (South Africa) No. 111749 (CPRR).
L.Y. acknowledges the Slovenian Research Agency grant number Z1-8139(B) for this work.
\footnotesize{
\bibliography{ref}
}

\end{document}